\documentclass{elsart}
\usepackage{amsfonts}
\usepackage{amssymb}
\usepackage{amsmath}

\input{tcilatex}

\begin{document}

\begin{frontmatter}

\title{Long Memory and Volatility Clustering: is the empirical evidence consistent across stock markets?}

\author{Sónia R. Bentes*}, {Rui Menezes**}, {Diana A. Mendes**}

\address{*Iscal, Av. Miguel Bombarda, 20, 1069-035 Lisboa Portugal, soniabentes@clix.pt;
 **ISCTE, Av. Forcas Armadas, 1649-025 Lisboa, Portugal }

\begin{abstract}
Long memory and volatility clustering are two stylized facts frequently related to financial markets. Traditionally, these phenomena have been studied based on conditionally heteroscedastic models like ARCH, GARCH, IGARCH and FIGARCH, inter alia. One advantage of these models is their ability to capture nonlinear dynamics. Another interesting manner to study the volatility phenomena is by using measures based on the concept of entropy. In this paper we investigate the long memory and volatility clustering for the SP 500, NASDAQ 100 and Stoxx 50 indexes in order to compare the US and European Markets. Additionally, we compare the results from conditionally heteroscedastic models with those from the entropy measures. In the latter, we examine Shannon entropy, Renyi entropy and Tsallis entropy. The results corroborate the previous evidence of nonlinear dynamics in the time series considered.
\end{abstract}


\end{frontmatter}

\section*{Introduction}

The study of stock market volatility and the reasons that lie beyond price
movements have always played a central role in financial theory, given rise
to an intense debate in which long memory and volatility clustering have
proven to be particularly significant. Since long memory reflects long run
dependencies between stock market returns, and volatility clustering
describes the tendency of large changes in asset prices to follow large
changes and small changes to follow small changes, these concepts are
interrelated and frequently studied in complementarity. Several models based
on heteroscedastic conditionally variance have been proposed to capture
their properties. This constitutes what we consider the traditional
approach. It includes the autoregressive conditionally heteroscedastic model
(ARCH) proposed by Engle \cite{Engle}, the Generalized ARCH (GARCH) due to
Bollerslev \cite{Bollerslev} and Taylor \cite{Taylor}, the Integrated GARCH
(IGARCH) derived by Engle and Bollerslev \cite{EngleBollerslev} and the
Fractionally Integrated GARCH (FIGARCH) introduced by Baillie \emph{et al.} 
\cite{Baillieetal}. These models account for nonlinear dynamics, which are
shown by the seasonal or cyclical behavior of many stock market returns and
constitute their main advantage. However, they are not fully satisfactory,
especially when modeling volatility of intra-daily financial returns. For a
comprehensive debate on this matter see Bordignon \emph{et al.} \cite%
{Bordignon}.

In this paper we propose an alternative way to study stock market volatility
based on econophysics models. The application of concepts of physics to
explain economic phenomena is relatively recent and started when some
regularities between economic/financial and physical data were found in a
consistent way (see e.g \cite{Stanley} and \cite{Plerou}). In this sense,
one concept of physics that can be helpful to measure the nonlinear
volatility of stock markets is the concept of entropy. Regarding this, we
discuss three different measures: Shannon entropy, Renyi entropy and Tsallis
entropy, and compare the main results.

The plan for the remainder of the paper is as follows: Section 2 puts
together the ARCH/GARCH type models and the entropy models. Next, in Section
3 we describe the empirical findings. Finally, Section 4 presents the
conclusions.

\section{Traditional Volatility Models versus Econophysics Models}

According to the traditional approach the presence of conditionally
heteroscedastic variance in stock market returns gives support to the use of
ARCH/GARCH models when studying stock market volatility. Although there are
other econometric models to seek for long memory and volatility clustering
the most common ones are the GARCH $(p,q)$ model and their derivations -
IGARCH $(p,q)$ and FIGARCH $(p,d,q)$ - which are summarized below.

Consider a time series $y_{t}$ with the associated error

\begin{equation}
e_{t}=y_{t}-E_{t-1}y_{t},
\end{equation}

where $E_{t-1}$ is the expectation operator conditioned on time $t-1$. A
GARCH model where%
\begin{equation}
e_{t}=z_{t}\sigma _{t},\qquad z_{t}\sim N\left( 0,1\right)
\end{equation}%
was developed such as

\begin{equation}
\sigma _{t}^{2}=\omega +\alpha \left( L\right) \varepsilon _{t}^{2}+\beta
\left( L\right) \sigma _{t}^{2},  \label{GARCH}
\end{equation}%
where $\omega >0$, and $\alpha \left( L\right) $ and $\beta \left( L\right) $
are polynomials in the lag operator $L$ $\left( L^{i}x_{i}=x_{t-i}\right) $
of order $q$ and $p,$ respectively. Expression (\ref{GARCH}) can be
rewritten as the infinite-order ARCH $(p)$ process,

\begin{equation}
\Phi \left( L\right) e_{t}^{2}=\omega +\left[ 1-\beta \left( L\right) \right]
\upsilon _{t},
\end{equation}

where $\upsilon _{t}\equiv e_{t}^{2}-\sigma _{t}^{2}$ and $\Phi \left(
L\right) =\left[ 1-\alpha \left( L\right) -\beta \left( L\right) \right] $.
Even though this process is frequently used to describe volatility
clustering, it shows some limitations when dealing with long memory since it
assumes that shocks decay at a fast geometric rate allowing only for short
term persistence. To overcome this drawback Engle and Bollerslev \cite%
{EngleBollerslev} developed the IGARCH specification given by 
\begin{equation}
\Phi \left( L\right) \left( 1-L\right) e_{t}^{2}=\omega +\left[ 1-\beta
\left( L\right) \right] \upsilon _{t}.
\end{equation}

Motivated by the presence of apparent long memory in the autocorrelations of
squared or absolute returns of several financial assets, Baillie \emph{et al.%
} \cite{Baillieetal} introduced the FIGARCH model defined as

\begin{equation}
\Phi \left( L\right) \left( 1-L\right) ^{d}e_{t}^{2}=\omega +\left[ 1-\beta
\left( L\right) \right] \upsilon _{t},
\end{equation}%
where $0\leq d\leq 1$ is the fractional difference parameter. An interesting
feature of this model is that it nests both the GARCH model for $d=0$ and
IGARCH for $d=1$. Alternatively, for $0<d<1$ the FIGARCH model implies a
long memory behavior, \emph{i.e.}, a slow decay of the impact of a
volatility shock. Also, we shall note that this type of processes is not
covariance stationary but instead strictly stationary and ergodic for $d\in %
\left[ 0,1\right] .$

An alternative way to study stock market volatility is by applying concepts
of physics which significant literature has already proven to be helpful in
describing financial or economic problems. One measure that can be applied
to describe the nonlinear dynamics of long memory and volatility clustering
is the concept of entropy. This concept was originally introduced in 1865 by
Clausius to explain the tendency of temperature, pressure, density and
chemical gradients to flatten out and gradually disappear over time. Based
on this Clausius developed the Second Law of Thermodynamics which postulates
that the entropy of an isolated system tends to increase continuously until
it reaches its equilibrium state. Although there are many different
understandings of this concept the most commonly used in literature is as a
measure of ignorance, disorder, uncertainty or even lack of information (see 
\cite{Golan}). Later, in a subsequent investigation Shannon \cite{Shannon}
provided a new insight on this matter showing that entropy wasn't only
restricted to thermodynamics but could be applied in any context where
probabilities can be defined. In fact, thermodynamic entropy can be viewed
as a special case of the Shannon entropy since it measures probabilities in
the full state space. Based on the Hartley's \cite{Hartley} formula, Shannon
derived his entropy measure and established the foundations of information
theory.

For the probability distribution $p_{i}\equiv p\left( X=i\right) $, $\left(
i=1,...,n\right) $ of a given random variable $X,$ Shannon entropy $S(X)$
for the discrete case, can be defined as 
\begin{equation}
S\left( X\right) =-\sum\limits_{i=1}^{n}p_{i}\ln p_{i},
\end{equation}%
with the conventions $0\ln \left( 0/z\right) =0$ for $z\geq 0$ and $z\ln
\left( z/0\right) =\infty $.

As a measure of uncertainty the properties of entropy are well established
in literature (see \cite{ShannonWeaver}). For the non-trivial case where the
probability of an event is less than one, the logarithm is negative and the
entropy has a positive sign. If the system only generates one event, there
is no uncertainty and the entropy is equal to zero. By the same token, as
the number of likely events duplicates the entropy increases one unit.
Similarly, it attains its maximum value when all likely events have the same
probability of occurrence. On the other hand, the entropy of a continuous
random variable may be negative. The scale of measurements sets an arbitrary
zero corresponding to a uniform distribution over a unit volume. A
distribution which is more confined than this has less entropy and will be
negative.

By replacing linear averaging in Shannon entropy with the Kolmogorov-Nagumo
average or quasi-linear mean and further imposing the additivity
constraints, Renyi \cite{Renyi} proposed the first formal generalization of
the Shannon entropy. The need for a new information measure was due to the
fact that there were a number of situations that couldn't be explained by
Shannon entropy. As Jizba and Arimitsu \cite{Jizbab} pointed out Shannon's
information measure represents mere idealized information appearing only in
situations when the storage capacity of a transmitting channel is finite.

Using this formalism Renyi \cite{Renyi} developed his information measure,
known as Renyi entropy or Renyi information measure of $\alpha $ order, $%
S_{\alpha }(X)$. For discrete variables it comes%
\begin{equation}
S_{\alpha }(X)=\frac{1}{1-\alpha }\ln \left(
\sum\limits_{i=1}^{n}p_{k}^{\alpha }\right) ,
\end{equation}%
\ 

for $\alpha >0$ and $\alpha \neq 1$. In the limit $\alpha \rightarrow 1$,
Renyi entropy reduces to Shannon entropy and can be viewed as a special case
of the latter. Additionally, evidence was found that Renyi's entropies of
order greater than 2 are related to search problems (see for example \cite%
{Pfister}). Even though this measure can be used in a variety of problems,
empirical evidence has shown that it has a built-in predisposition to
account for self-similar systems and, so, it naturally aspires to be an
effective tool to describe equilibrium and non-equilibrium phase transitions
(see \cite{Jizbaa} and \cite{Jizbab}). Despite its relevance, Renyi entropy
didn't experience the same success of its predecessor's which can be
explained basically by two factors: ambiguous renormalization of Renyi's
entropy for non-discrete distributions and little insight into the meaning
of Renyi's $\alpha $ index (see \cite{Jizbaa}). A new insight into this
matter was brought by Csisz\'{a}r \cite{Csiszar}, who has identified the $%
\alpha $ index as the $\beta $ cutoff rate for hypothesis testing problems.

With the aim of studying physical systems that entail long-range
interaction, long-term memories and multi-fractal structures, Tsallis \cite%
{Tsallis} derived a new generalized form of entropy, known as Tsallis
entropy. Although this measure was first introduced by Havrda and Charv\'{a}%
t \cite{HavdraCharvat} in cybernetics and late improved by Dar\'{o}czy \cite%
{Daroczy}, it was Tsallis \cite{Tsallis} who really developed it in the
context of physical statistics and, therefore, it is also known as
Havrda-Charv\'{a}t-Dar\'{o}czy-Tsallis entropy.

For any nonnegative real number $q$ and considering the probability
distribution $p_{i}\equiv p$ $(X=i)$, $i=1,...,n$ of a given random variable 
$X,$ Tsallis entropy denoted by $S_{q}\left( X\right) $, is defined as 
\begin{equation}
S_{q}\left( X\right) =\frac{1-\sum\limits_{i=1}^{n}p_{i}^{q}}{q-1}.
\end{equation}

As $q\rightarrow 1,$ $S_{q}$ recovers $S_{q}\left( X\right) $ because the $q$%
-logarithm uniformly converges to a natural logarithm as $q\rightarrow 1$.
This index may be thought as a biasing parameter since $q<1$ privileges rare
events and $q>1$ privileges common events (see \cite{Tsallisetal}). A
concrete consequence of this is that while Shannon entropy yields
exponential equilibrium distributions, Tsallis entropy yields power-law
distributions. As Tatsuaki and Takeshi \cite{Tatsuaki} have already pointed
out the index $q$ plays a similar role as the light velocity $c$ in special
relativity or Planck's constant $\hbar $ in quantum mechanics in the sense
of a one-parameter extension of classical mechanics, but unlike $c$ or $%
\hbar $, $q$ does not seem to be a universal constant. Further, we shall
also mention that for applications of finite variance $q$ must lie within
the range $1\leq q<5/3$.

\section{Empirical evidence}

This section examines the results obtained from both perspectives. In order
to compare the volatility of the US and European stock market returns we
have collected data from SP 500, NASDAQ 100 and Stoxx 50 indexes and
constituted a sample spanning over the period June 2002-January 2007. The
values were gathered on a daily basis without considering the re-investment
of dividends. Based on them, we computed the stock market returns given by
the log-ratio of the index values at time $t$ and time $t-1$ and performed
the estimates.

Within the traditional approach we have considered the GARCH $\left(
1,1\right) $, IGARCH $\left( 1,1\right) $ and FIGARCH $\left( 1,d,1\right) $
specifications, whose main results are listed in Table \ref{ARCH}. The
conclusions are similar to all the three indexes considered. Specifically,
for the GARCH $\left( 1,1\right) $ it was found evidence of heteroscedastic
conditional variance. Also, the fact that $\alpha +\beta \simeq 1$ could
denounce the presence of nonlinear persistence in the log-returns of the
stock market indexes led us to estimate the IGARCH $(1,1)$ model. However,
evidence has shown that many coefficients were not statistically
significant. Then, the next step was to adjust the FIGARCH specification $%
(1,d,1)$ with the restriction $d\neq 1$ whose main results corroborate the
long memory hypothesis.

\begin{table}[h]
\begin{tabular}{ccccc}
\hline\hline
{\small Coef.} & Indexes & GARCH & IGARCH & FIGARCH \\ \hline
& \multicolumn{1}{l}{Stoxx 50} & $1.12E-06$* & $0.014544$ & $6.573658$** \\ 
$\omega $ & \multicolumn{1}{l}{SP 500} & $3.19E-07$** & $0.003906$* & $%
0.009144$ \\ 
& \multicolumn{1}{l}{NASDAQ 100} & $5.74e-07$* & $0.004543$ & $8.749987$**
\\ \hline
& \multicolumn{1}{l}{Stoxx 50} & $0.076581$** & $0.104788$** & $-0.119727$**
\\ 
$\alpha $ & \multicolumn{1}{l}{SP 500} & $0.051592$** & $0.057835$** & $%
-0.118990$ \\ 
& \multicolumn{1}{l}{NASDAQ 100} & $0.040982$** & $0.044868$** & $-0.037189$%
** \\ \hline
& \multicolumn{1}{l}{Stoxx 50} & $0.913627$** & $0.895212$ & $0.643566$** \\ 
$\beta $ & \multicolumn{1}{l}{SP 500} & $0.946445$** & $0.942165$ & $%
0.514637 $* \\ 
& \multicolumn{1}{l}{NASDAQ 100} & $0.957592$** & $0.955132$ & $0.700161$**
\\ \hline
& \multicolumn{1}{l}{Stoxx 50} & - & - & $0.695285$** \\ 
$d$ & \multicolumn{1}{l}{SP 500} & - & - & $0.579649$** \\ 
& \multicolumn{1}{l}{NASDAQ 100} & - & - & $0.655988$** \\ \hline
& \multicolumn{1}{l}{Stoxx 50} & $10.15165$** & $25.044983$ & $21.850493$*
\\ 
Student & \multicolumn{1}{l}{SP 500} & $6.152842$** & $5.374068$** & $%
1219.021742$** \\ 
& \multicolumn{1}{l}{NASDAQ 100} & $11.42494$** & $11.030880$** & $%
335.539275 $** \\ \hline
& \multicolumn{1}{l}{Stoxx 50} & $3812.074$ & $1290.958$ & $1296.348$ \\ 
Log-L & \multicolumn{1}{l}{SP 500} & $13891.33$ & $13039.1$ & $1482.22$ \\ 
& \multicolumn{1}{l}{NASDAQ 100} & $10775.58$ & $10024.569$ & $1248.274$ \\ 
\hline\hline
\end{tabular}%
\caption{{\protect\small GARCH, IGARCH and FIGARCH models for Stoxx 50, SP
500 and NASDAQ 100 indexes; ** denotes significance at the 1\% level, *
denotes significance at the 5\% level}}
\label{ARCH}
\end{table}

\bigskip In the domain of the econophysics approach we have computed the
Shannon, Renyi and Tsallis entropies which are depicted in Table \ref%
{entropias}. 
\begin{table}[h]
\begin{tabular}{ccccc}
\hline\hline
{\small Entropies} & Index ($\alpha $/$q$) & Stoxx & SP 500 & NASDAQ 100 \\ 
\hline
Shannon & - & $3.3624$ & $3.3784$ & $3.2981$ \\ \hline
& $1.4$ & $10.2076$ & $10.217$ & $10.2085$ \\ 
Renyi & $1.45$ & $10.2065$ & $10.2163$ & $10.2074$ \\ 
& $1.5$ & $10.2054$ & $10.2155$ & $10.2064$ \\ \hline
& $1.4$ & $1.8354$ & $1.8395$ & $1.8102$ \\ 
Tsallis & $1.45$ & $1.7204$ & $1.7238$ & $1.6981$ \\ 
& $1.5$ & $1.6161$ & $1.619$ & $1.5965$ \\ \hline\hline
\end{tabular}%
\caption{{\protect\small Shannon, Renyi and Tsallis entropies}}
\label{entropias}
\end{table}

All entropies were estimated with histograms based on equidistant cells. For
the calculation of Tsallis entropy we set values at $1.4$, $1.45$ and $1.5$
for the index $q$, which is consistent with the finding that when
considering financial data their values lie within the range $q\simeq
1.4-1.5 $ (see \cite{Tsallisetal}). The same assumption was made for the
Renyi's index. Since all entropies are positive we shall conclude that the
data show nonlinearities. This phenomenon is particularly evident for the SP
500 index, which always attained the highest levels regardless of the method
applied in its calculation. As for the others the results are not conclusive
since they vary according to the entropy method adopted.

\section{Conclusions}

In this paper we have investigated the properties of the realized volatility
for the SP 500, NASDAQ 100 and Stoxx 50 indexes. Our main goal was to
compare two different perspectives: the so-called traditional approach in
which we have considered the GARCH $\left( 1,1\right) $, IGARCH $\left(
1,1\right) $ and FIGARCH $\left( 1,d,1\right) $ specifications and the
econophysics approach based on the concept of entropy. For our purpose three
variants of this notion were chosen: the Shannon, Renyi and Tsallis
measures. The results from both perspectives have shown nonlinear dynamics
in the volatility of SP 500, NASDAQ 100 and Stoxx 50 indexes and must be
understood in complementarity.

We consider that the concept of entropy can be of great help when analyzing
stock market returns since it can capture the uncertainty and disorder of
the time series without imposing any constraints on the theoretical
probability distribution. By contrast, the ARCH/GARCH type models assume
that all variables are independent and identically distributed (\emph{i.i.d}%
). However, in order to capture global serial dependence one should use a
specific measure such as, for example, mutual information. By analyzing the
entropy values for different equally spaced sub-periods we could have a
clearer idea about the extent of volatility clustering and long-memory
effects, an issue that will be pursued in further work.

\end{document}